\documentclass[twocolumn,prb,showpacs,preprintnumbers,amsmath,amssymb]{revtex4}

\usepackage{graphicx}
\usepackage{dcolumn}
\usepackage{bm}

\begin{document}


\title{Fluctuation diamagnetism around the superconducting transition in a cuprate crystal with a reduced Meissner fraction}


\author{Jes\'us Mosqueira}
\author{F\'elix Vidal}
\affiliation{LBTS, Facultad de F\'isica, Universidade de Santiago de Compostela, E-15782 Spain}

\date{\today}

\begin{abstract}
The magnetization around the superconducting transition was measured in a Tl$_{0.5}$Pb$_{0.5}$Sr$_2$CaCu$_2$O$_7$ crystal affected by a considerable reduction ($\sim$55\%) of its effective superconducting volume fraction but still with a relatively sharp low-field Meissner transition, a behaviour  that may be attributed to the presence of structural inhomogeneities. By taking into account these inhomogeneities just through the Meissner fraction, the observed diamagnetism may still be explained, consistently above and below the superconducting transition, in terms of the conventional Ginzburg-Landau approach with fluctuations of Cooper pairs and vortices.
\end{abstract}

\maketitle

The behavior of the magnetization around the Meissner transition provides an unavoidable constraint for any phenomenological description of a superconducting transition.\cite{Tinkham} In the last few years, various groups have reported the observation of \textit{strong anomalies} in the magnetization measured around the superconducting transition in high-$T_C$ cuprate superconductors (HTSC) with different doping levels. Among these anomalies are the observation under low fields of \textit{giant} diamagnetism (with amplitudes orders of magnitude larger than the one associated with superconducting fluctuations in the conventional Ginzburg-Landau (GL) scenario) and a seemingly non-linear temperature behavior of the associated upper critical field, $H_{C2}(T)$, near $T_C$.\cite{Carretta, Wang,Cabo} The origin of this unconventional (non-GL) behavior is at present a debated issue,\cite{Carretta, Wang,Cabo,Ovchinnikov,Sewer,deMello,Alex06,Oganesian,Ander1,Comment,Reply,Doria,Podolsky} the proposals including $T_C$ inhomogeneities or vortex fluctuations even well above the measured $T_C$. The interest of this debate is enhanced by the fact that it also concerns other open aspects of the HTSC, as the pseudogap in the normal state or the possible existence of a vortex fluid over a wide temperature range above $T_C$.\cite{Podolsky,Lee,Ander2}
 
The magnetization measurements and analyses performed recently in our group in different HTSC and \textit{dirty} low $T_C$ superconductors (without nonlocal electrodynamic effects) favours the presence of extrinsic $T_C$ inhomogeneities, just associated with chemical inhomogeneities, as the origin of most of the observed magnetization  anomalies.\cite{Cabo,Comment,Tl2223,Nota1}. Nevertheless, there is another very common type of inhomogeneity whose influence on the magnetization also deserves a close inspection: the one associated with structural defects at different length scales, including those as smaller as a few times the superconducting coherence length amplitude, $\xi(0)$. In extreme type II superconductors, even these short length inhomogeneities, difficult to be directly observed, may strongly decrease the effective superconducting volume fraction without enlarging the temperature width of the low-field Meissner transition. In this Brief Report, we will first present detailed magnetization measurements around the Meissner transition in a Tl$_{0.5}$Pb$_{0.5}$Sr$_2$CaCu$_2$O$_7$ (TlPb1212) crystal deeply affected by a reduction of the effective volume fraction, an effect which does not enlarge the temperature width of the low-field Meissner transition and that may be attributed to structural inhomogeneities. Then, it will be shown that if these anomalies are taken into account through the Meissner fraction, the diamagnetism around $T_C$ may still be explained in terms of the conventional Ginzburg-Landau approach with fluctuations of Cooper pairs and vortices. This agreement extend to all the different fluctuation regions in the $H-T$  phase diagram, thus generalizing previous results  for the so-called \textit{crossing-point} of the magnetization versus temperature curves.\cite{volume_fraction}

The TlPb-1212 sample used in this work is a $1.10\times0.85\times0.192$ mm$^3$ single crystal. Details of its growth procedure and subsequent structural characterization may be found in Ref.~\onlinecite{Maignan}. Let us only mention that x-ray diffraction revealed that it was single phase, with a well defined $c$-crystallographic length of $c=12.1$ \r{A}. The magnetization measurements were performed with a superconducting-quantum-interference-device (SQUID) magnetometer (Quantum Design). As a first magnetic characterization, we measured the temperature dependence of the field-cooled (FC) magnetic susceptibility with a 1 mT magnetic field applied perpendicularly to the $ab$ crystallographic planes. The result is presented in the lower inset of Fig.~1, already corrected for demagnetizing effects. For that, we used the demagnetizing factor resulting from the sample dimensions by using the ellipsoidal approximation. For our sample this leads to a factor of $D=0.75$. These last data show that this crystal has a narrow diamagnetic transition, the relative width being $\Delta T_C/T_C\sim3\times10^{-2}$ with the midpoint at $T_C=77.1$ K. However, they also reveal a strong reduction (around 55\%) of its effective superconducting volume fraction. These results provide a quite direct indication that this crystal is deeply affected by (temperature independent) structural inhomogeneities uniformly distributed in the sample volume which, as noted before, are particularly effective in reducing the Meissner fraction of the extreme type II superconductor studied here.

The excess diamagnetism for $H\perp ab$ was obtained by subtracting to the raw data the normal-state contribution, determined by fitting a Curie-like function to the $M(T,H)$ curves well above $T_C$ (between $\sim$100 K and $\sim$200 K). An overview of the resulting $\Delta M(T)$ in all the reversible region is presented in Fig.~1 for magnetic fields between 0.5 T and 5 T . As may be seen in the upper inset, the curves for $\mu_0H\geq 2$ T cross at a temperature $T^*_1\approx75$ K, which is a signature of the thermal fluctuations in highly anisotropic superconductors. For lower field amplitudes, the crossing point shifts to a higher temperature ($T_2^*\approx 76$ K). This behavior, which may be seen more clearly in the $\Delta M(H)_T$ representation (see below), was already observed in other highly anisotropic superconductors,\cite{Naughton,Tl2223} and may be attributed to a change of the fluctuation regime induced by the magnetic field. 

\begin{figure}[b]
\includegraphics[scale=0.5]{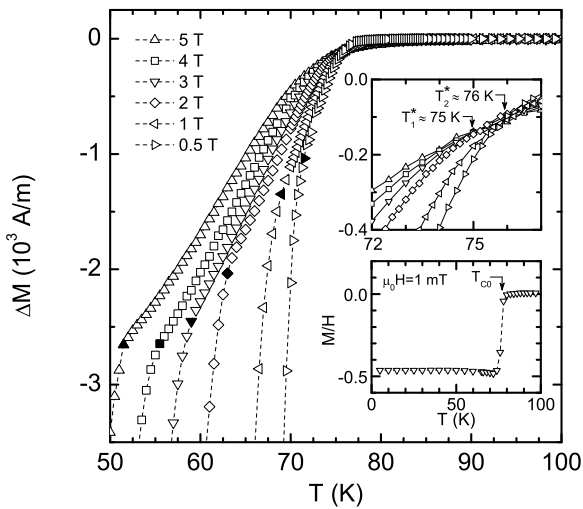}
\caption{Overview of the $T$ dependence of the fluctuation magnetization for $H\perp ab$ in the reversible region around $T_C$. The solid symbols indicate the transition to the irreversible mixed state. Upper inset: Detail of the crossing point which reveals its splitting: the $\Delta M(T)$ curves for $\mu_0H\geq2$ T cross at $T_1^*$, while for lower fields cross at $T_2^*$. Lower inset: $T$ dependence of the low-field FC magnetic susceptibility, already corrected for demagnetizing effects.}
\end{figure}

For temperatures and magnetic fields well above $T_C(H)$, the fluctuation magnetization of highly anisotropic superconductors when $H\perp ab$ predicted by the GL theory in the Gaussian approximation (GGL approach) and taking into account the total energy cutoff reads,\cite{Tl2223,EPLYBCO,EPLvidal,PCCarballeira}
\begin{eqnarray}
\Delta M=-f\frac{k_BTN}{\phi_0s}\left[-\frac{\varepsilon^c}{2h}\psi\left(\frac{h+\varepsilon^c}{2h}\right)-\ln\Gamma\left(\frac{h+\varepsilon}{2h}\right)\right.\nonumber \\
+\left.\ln\Gamma\left(\frac{h+\varepsilon^c}{2h}\right)+\frac{\varepsilon}{2h}\psi\left(\frac{h+\varepsilon}{2h}\right)+\frac{\varepsilon^c-\varepsilon}{2h}\right].
\label{Prange}
\end{eqnarray}
where the notation is the same as for Eq.~(1) in Ref. \onlinecite{Tl2223}. Some examples of the measured $\Delta M$ above $T_{C0}$ are presented in Fig.~2. The lines correspond to Eq.~(\ref{Prange}) with $N=2$, $s=c=12.1$ \r{A}, $\varepsilon^c=0.55$, $f$ approximated by the Meissner fraction $|\chi^{FC}(0)|\approx0.45$, and $H_{C2}(0)$ as the only free parameter. As may be clearly seen, the agreement with the experimental data is excellent down to few degrees above $T_{C0}$, where the Gaussian approximation is no longer valid, and it leads to $\mu_0H_{C2}(0)\approx170$ T, a value that is going to be used in the remaining analyses. Note that Eq.~(\ref{Prange}) is also in excellent agreement with the $H$ dependence of $\Delta M$ (inset of Fig.~2). This behavior is expected to hold up to $h\sim0.1$,\cite{PRLCarballeira} where fluctuation effects begin to decrease due to quantum effects associated with the shrinkage of the superconducting wavefunction.\cite{EPLvidal,Soto} If an ideal superconducting volume fraction is assumed ($f=1$), the agreement is also good, but leads to a different $\mu_0H_{C2}(0)$ value ($\sim 370$ T), which has consequences in the subsequent analysis. 

\begin{figure}[t]
\includegraphics[scale=0.5]{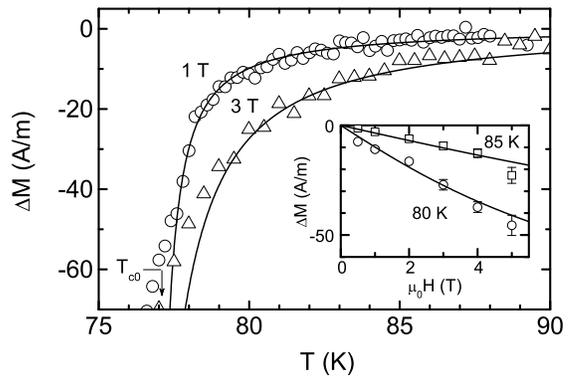}
\caption{Fluctuation magnetization vs. $T$ and vs. $H$ (inset) in the Gaussian region above $T_C(H)$. The lines are the GGL result for finite $H$ [Eq.(\ref{Prange})]. }
\end{figure}

\begin{figure}[b]
\includegraphics[scale=0.5]{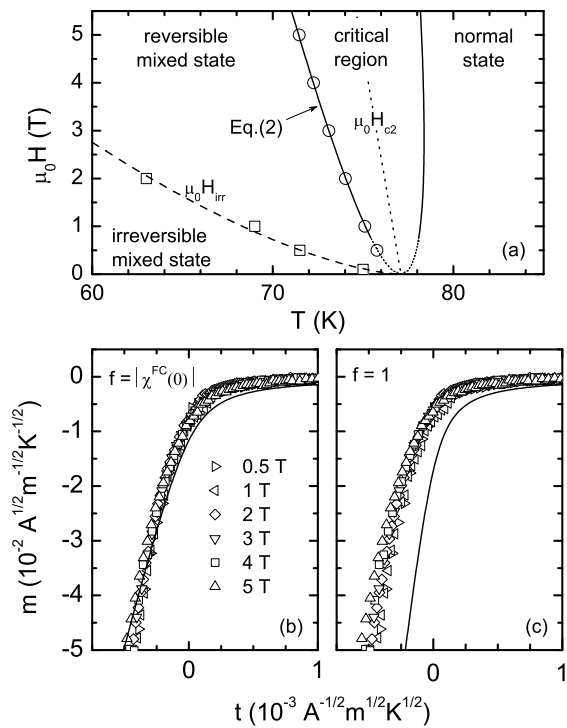}
\caption{a) $H-T$ superconducting phase diagram indicating the different fluctuation regions around $H_{C2}(T)$. Circles are the low-temperature limit of the region where the scaling approach holds, and the solid line is the best fit of the $H$-dependent Ginzburg criterion. Squares are the irreversibility line as deduced from Fig.~1, and the dashed line a fit to a $(T_C-T)^{3/2}$ power law. b) Scaling of the $m$ vs $t$ curves in the critical region around $H_{C2}(T)$. The line is the GL-LLL scaling function [Eq.~(\ref{scalingfunction})]. In obtaining the scaling variables and function, we used the $H_{C2}(0)$ value resulting from the $\Delta M$ analysis in the Gaussian region well above $T_C$ (Fig.~2). c) The same analysis assuming an ideal effective superconducting volume fraction ($f=1$).}
\end{figure}

For temperatures closer to $T_C(H)$, in the so-called critical region [see Fig.~3(a)], the Gaussian approximation is no longer valid. In a magnetic field sufficiently strong that the Cooper pairs are limited to the lowest Landau-level, this critical region is bounded by the so-called field-dependent Ginzburg criterion,\cite{Ikeda} which for two-dimensional systems is given by \cite{Kim}
\begin{equation}
|T-T_C(H)|/T_{C0}\stackrel{<}{_\sim}\sqrt{4\pi k_B\mu_0H/\phi_0s\Delta c},
\label{criterio}
\end{equation}
where $\Delta c$ is the specific heat jump at $T_{C0}$. In this regime the GL theory predicts that the fluctuation induced magnetization follow a scaling behavior in the variables\cite{Ullah}
\begin{equation}
m\equiv \Delta M/\sqrt{HT},\;\;\;t\equiv [T-T_C(H)]/\sqrt{HT}.
\label{scaling}
\end{equation}
By using a non-perturbative approach to the GL free energy in the lowest Landau-level approximation (GL-LLL approach), Te\u{s}anovi\'{c} \textit{et al.} obtained an explicit equation for the scaling function, which may be written as \cite{Tesanovic}
\begin{equation}
m=f\frac{A}{H_{C2}^{'}}\frac{k_B}{\phi_0s}\left(At-\sqrt{A^2t^2+2}\right),
\label{scalingfunction}
\end{equation}
where $A\equiv[H_{C2}^{'}T^*_1/2(T_{C0}-T^*_1)]^{1/2}$, $H_{C2}^{'}\equiv H_{C2}(0)/T_{C0}$, and $T^*_1$ corresponds to the limit of the critical region below $T_{C0}$ when $H=0$. This expression predicts the crossing of the $M(T)_H$ curves at $T^*_1$, and gives for the crossing point magnetization
\begin{equation}
\Delta M^*_1=-f\frac{k_BT^*_1}{\phi_0s}.
\label{crucealto}
\end{equation}
This equation allows a direct comparison with the experiments. As may be easily checked, the high-field crossing point observed at $T_1^*$ falls into the critical region bounded by Eq.~(\ref{criterio}) and should be described by Eq.~(\ref{crucealto}). By using $f=|\chi^{FC}(0)|$, it leads to $\Delta M^*_1\approx-190$ A/m, in relative good agreement with the experimental value ($-150$ A/m) taking into account the experimental uncertainties in $\chi^{FC}(0)$ and in the normal-state $M_B(T)$ contribution. A similar agreement was also found in a variety of highly anisotropic HTSC with different $\chi^{FC}(0)$ values.\cite{volume_fraction} However, by imposing $f=1$ as would correspond to an ideal sample, the disagreement is well beyond these uncertainties. 

In Fig.~3(b) we present the scaling of the $\Delta M(T)_H$ data in the critical region according to Eqs.~(\ref{scaling}). The scaling variable $t$ is obtained by assuming a linear $H$ dependence of the critical temperature, $T_C(H)=T_{C0}[1-H/H_{C2}(0)]$, and by using $\mu_0H_{C2}(0)=170$ T, as results from previous analysis in the Gaussian region above $T_C$. The line in this figure is the scaling function [Eq.~(\ref{scalingfunction})] calculated with the same $H_{C2}(0)$ value. As may be clearly seen, the scaling of the $m(t)$ curves is excellent 
and the scaling function is also in good agreement with the data. The low-temperature limit of the region where the scaling holds is represented as circles in the $H-T$ phase diagram of Fig.~3(a). The fit of Eq.~(2) to these data (solid line) is excellent and leads to $\Delta c\approx 1.7\times10^5$ J/Km$^3$, which is close to the value found in other highly anisotropic HTSC.\cite{deltaC} For completeness, in Fig.~3(c) we present the $m(t)$ data evaluated by using $\mu_0H=370$ T (the value resulting from previous analysis if $f=1$ is imposed). As may be clearly seen, the scaling is considerably worsened, and also the scaling function is far from the data points.

For temperatures well below $T_C(H)$, outside the critical region delimited by Eq.~(\ref{criterio}), the fluctuations of the order parameter amplitude are negligible. However, the highly anisotropic nature of this compound lead to a contribution to the magnetization associated to thermal fluctuations of the two-dimensional vortex positions. This contribution has been calculated by Bulaevskii, Ledvig and Kogan in the framework of the GL theory.\cite{BLK} For $H\perp ab$ it may be expressed as
\begin{eqnarray}
\Delta M(T,H)=-f\frac{\phi_0}{8\pi\mu_0\lambda^2_{ab}(T)}{\rm ln}\left(\frac{\eta H_{C2}(T)}{H}\right)+\nonumber\\
+f\frac{k_BT}{\phi_0s}{\rm ln}\left(\frac{8\pi\mu_0k_BT\lambda_{ab}^2(T)}{\alpha s\phi_0^2\sqrt{e}}\frac{H_{C2}(T)}{H}\right).
\label{BLK}
\end{eqnarray}
The first term on the right is the conventional London magnetization, whereas the second one is associated with vortex fluctuations. In this equation, $\lambda_{ab}$ is the magnetic penetration length in the $ab$ planes, and $\eta$ and $\alpha$ are constants around the unity. This equation also predicts the crossing of the $\Delta M(T)_H$ curves at a temperature $T_2^*<T_{C0}$, the magnetization at the crossing point being
\begin{equation}
\Delta M^*_2=-f\frac{k_BT^*_2}{\phi_0s}\ln{(\eta\alpha\sqrt{e})}.
\end{equation}
This expression is analogous to Eq.~(5) for the crossing point in the critical region, except for a constant around the unity. By combining both expressions we obtained $\ln(\eta\alpha\sqrt{e})=\Delta M^*_2T^*_1/\Delta M^*_1T^*_2\approx0.73$, as expected. The comparison of Eq.~(\ref{BLK}) with the experimental data is presented in Fig.~4 where, for convenience, $\Delta M$ is represented against the magnetic field for several constant temperatures. For each isotherm, the only free parameters are $\eta H_{C2}$ and $\lambda_{ab}$. The fit quality is excellent for isotherms up to $\sim T_2^*$, except for data under high fields [$H\sim H_{C2}(T)$] which are already inside the critical region. The resulting $\eta \mu_0H_{C2}(T)$ and $\lambda_{ab}^{-2}(T)$ are presented in the inset. They follow the GL linear temperature dependence, tending to zero at $T_{C0}\approx 77$ K in agreement with the precedent analysis. It is worth noting that the London theory alone fits the data of Fig.~4 as well (it follows the same $H$ dependence). However, it would lead to an anomalous temperature dependence of $H_{C2}$ and $\lambda_{ab}$, mainly close to $T_2^*$.\cite{Kogan} The same happens if an ideal effective volume fraction ($f=1$) is assumed.

\begin{figure}[t]
\includegraphics[scale=0.5]{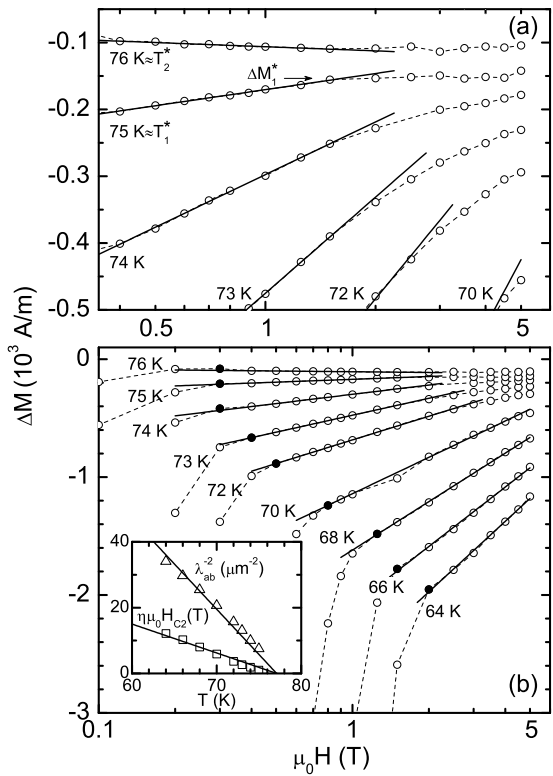}
\caption{$\Delta M$ vs $H$ in the Gaussian region below $H_{C2}(T)$. a) Detail around the crossing point temperatures. b) Overview of the reversible mixed state (solid symbols indicate the transition to the irreversible region). The lines are fits of the BLK theory [Eq.(\ref{BLK})], with $\lambda_{ab}$ and $\eta H_{C2}$ as the free parameters. The resulting $\eta\mu_0 H_{C2}$ and $\lambda_{ab}^{-2}$ (inset) follow the GL prediction (solid lines). See main text for details.}
\end{figure}

Summarizing, the diamagnetism anomalies observed around the Meissner transition in a cuprate superconductor deeply affected by a low effective superconducting  volume fraction may be easily overcome by just normalizing the magnetization through the low-field Meissner fraction: The resulting diamagnetism on both sides of $T_C$ may be explained in terms of the conventional Ginzburg-Landau approach with fluctuations of Cooper pairs and vortices. A remarkable new result of our present work when compared with previous magnetization measurements in other cuprate single crystals is the unambiguous demonstration of the need of a normalization through the Meissner fraction to eliminate the temperature independent anomalies in the diamagnetism amplitude.\cite{Nota2} Our results provide then a further confirmation that the Meissner transition in cuprate superconductors is a conventional GL transition, although in some cases entangled with chemical, structural or electronic inhomogeneities and disorder.  

We thank A. Maignan and A. Wahl for providing us the crystal, the Spanish Ministerio de \mbox{Educaci\'on} y Ciencia (Grant No. FIS2007-63709) and the Xunta de Galicia (Grant No. PGIDIT04TMT206002PR).


\begin{references}
\bibitem{Nota2}For instance, the single crystal measured in Ref. \onlinecite{Tl2223} has an effective superconducting volume fraction close to the ideal one, which makes more ambiguous the role of temperature independent inhomogeneities. Other crystals may have a low Meissner fraction but often this is entangled with $T_C$ inhomogeneities associated with chemical inhomogeneities (see, e.g., Refs.\onlinecite{Cabo,Nota1}).
\end{references}
\end{document}